\documentclass[article]{JHEP3}

\usepackage{amsmath,amssymb}
\usepackage[dvips]{graphics}
\usepackage{epsfig}

\newcommand{\be}{\begin{equation}}
\newcommand{\ee}{\end{equation}}
\newcommand{\bea}{\begin{eqnarray}}
\newcommand{\eea}{\end{eqnarray}}
\newcommand{\bean}{\begin{eqnarray*}}
\newcommand{\eean}{\end{eqnarray*}}

\relax

\def\d{\partial}
\def\a{\alpha'}
\def\R{\mathcal{R}}

\title{Tensorial perturbations and stability of spherically symmetric $d$--dimensional black holes in string theory}

\author{Filipe Moura
\\
Centro de Matem\'atica da Universidade do Minho, \\Escola de Ci\^encias, Campus de Gualtar, \\4710-057 Braga, Portugal\\
\\
\email{fmoura@math.uminho.pt}
}

\abstract{We compute the tensorial perturbations to a general spherically symmetric metric in $d$ dimensions with string--theoretical corrections quadratic in the Riemann tensor, from which we derive their respective potential. We use this result to study the stability of corresponding black hole solutions under such perturbations.
}

\begin{document}



\vfill

\eject

\section{Introduction and Summary}
\indent

In string theory there is a large variety of black hole solutions in four and more spacetime dimensions. Studying the stability of such exact solutions is a very important subject: if a stationary black hole solution is stable under perturbations, it implies that such solution describes a possible final state of dynamical evolution of a gravitating system. If, on the other hand, it is shown to be unstable, that indicates the existence of a different branch of solutions which the original solution may decay into, which means that one can anticipate a wider variety of black hole solutions. The analysis of perturbations also gives information about physical properties of the black hole solutions, an example being the spectra of quasinormal modes. It is interesting to study how such properties, including the stability, are affected in the context of string theory.

The analysis of linear perturbations of stationary black holes in arbitrary $d$ spacetime dimensions was initiated in \cite{iks00, ik03a, ik03c} for static, nonrotating black holes (without and with charge), for the three kinds of metric perturbations: tensor, vector and scalar. The analysis of stability was performed, and stability was obtained for asymptotically flat \cite{ik03b, ik03d} and anti--de Sitter black holes \cite{Konoplya:2008rq}, while an instability was found in de Sitter, for large charge and $\Lambda$ \cite{Konoplya:2008au}.

More recently the stability analysis was performed for rotating Myers--Perry black holes, first for tensor perturbations, where an instability was found in anti--de Sitter \cite{Kunduri:2006qa, Kodama:2009rq}, but not for asymptotically flat black holes. Considering time--dependent perturbations, for the first time an instability was found for asymptotically flat black holes, for sufficiently rapid rotations (the ultraspinning instability) \cite{Dias:2009iu, Dias:2010eu, Dias:2010maa}. Later the result was generalized for the case of anti--de Sitter \cite{Dias:2010gk}.

The analysis of perturbations and stability was also extended to Lovelock theories in $d$ dimensions, considering corrections quadratic \cite{dg04,dg05a} and cubic \cite{Takahashi:2009dz, Takahashi:2010ye, Takahashi:2010gz} in the Riemann tensor. Detailed studies of black hole perturbations in Lovelock theories were
performed, in the context of the AdS/CFT correspondence, in \cite{Camanho:2009vw,Camanho:2009hu,Camanho:2010ru}. Depending on the gravitational potential resulting from the couplings of  the theory, different kinds of instabilities were found, which could imply either causality violations or plasma instabilities in the holographic dual thermal field theory.

Although superstring theories also require higher order corrections to the Einstein--Hilbert action (actually to supergravity, which is the low energy effective theory coming from superstrings), there are important differences between string and Lovelock theory effective actions. In Lovelock theories, the actions are made just with powers of the Riemann curvature tensor (i.e. one can have an action just with the metric field and its derivatives); in string theory, besides the graviton we always have to consider at least the dilaton field which, as we will see later in this article (after eq. (\ref{bdfe})), cannot be set to zero in the presence of higher order terms. Besides, Lovelock actions are not seen as ``effective'' in the sense of string perturbation theory, but as ``exact''. To be precise, in perturbative string theory one only works up to the order of the perturbation constant (in the case of higher order string corrections, this is the inverse string tension $\a$) which appears in the effective action, neglecting all the higher order terms which are meaningless in such effective action, while in Lovelock theories one considers exact solutions to the equations of motion, no matter the order of the constant which appears in such solutions, even if the constant only appears to first order in the lagrangian (multiplying the higher order term) and in the field equations. Because of these differences, results and conclusions which are obtained in Lovelock theories may be very different from those in string theory.

String theory is the most promising candidate for a consistent description of quantum gravity; therefore, it is natural to study its effects on the stability of its black hole solutions. The study of tensorial perturbations to the metric and the respective stability of black holes with higher order corrections in the context of string theory was initiated in \cite{Moura:2006pz}, concerning a particular spherically symmetric solution. In this article we extend such study to incorporate the most general spherically symmetric black hole solutions with gravitational corrections to first order in $\a$ in string theory in arbitrary $d$ dimensions.

The article is organized as follows: in section 2, we review the general formalism for field perturbations, and then we compute the tensorial perturbation equations for the most general spherically symmetric background metric in $d$ dimensions. In section 3, we apply these perturbation equations to the field equations obtained from a string theory effective action with string corrections to first order in $\a$, i.e. quadratic in the Riemann tensor. We obtain the master equation for the perturbation variable and the potential for tensorial perturbations including such corrections. Then in section 4 we review the ``S--deformation'' approach and we obtain a criterion for the gravitational stability of black holes under tensor perturbations from the master equation. Finally on section 5 we apply this criterion to study the stability of two different solutions with leading $\a$ corrections in $d$ dimensions: the dilatonic compactified black hole and the double--charged black hole.


\section{General setup of the perturbation theory}

\subsection{Perturbations on a $(d-2)$--sphere}
\indent

We will study the behavior, under gravitational perturbations, of string--corrected black hole solutions in a generic spacetime dimension $d$. For such analysis we use the framework developed by Ishibashi and Kodama \cite{iks00,ik03a,ik03c} for black holes. This framework applies to generic spacetimes of the form $\mathcal{M}^{d} = \mathcal{N}^{d-n} \times \mathcal{K}^n$, with coordinates $\left\{ x^\mu \right\} = \left\{ y^a, \theta^i \right\}$. In here $\mathcal{K}^n$ is a manifold with constant sectional curvature $K.$ The metric in the total space $\mathcal{M}^{d}$ is then written as

\be \label{ikmetric}
g = g_{ab}(y)\ dy^a\ dy^b + r^2(y)\ \gamma_{ij}(\theta)\ d\theta^i\ d \theta^j.
\ee
For our purposes, we take $n = d-2$ and the manifold $\mathcal{K}^n$, describing the geometry of the black hole event horizon, will be a $(d-2)$--sphere (thus, with $K=1$). Also, $\mathcal{N}^{d-n}$ coordinates will be $\left\{ y^a \right\} = \left\{ t, r \right\},$ with $\left\{ r, \theta^i \right\}$ being the usual spherical coordinates so that $r(y) = r$ and $\gamma_{ij}(\theta)\ d\theta^i d\theta^j = d\Omega_{d-2}^2$.

Defining generic perturbations to the metric as $h_{\mu\nu} = \delta g_{\mu\nu}, \, h^{\mu\nu} = -\delta g^{\mu\nu},$
we get for the variation of the Levi-Civita connection

\be \label{deltagamma}
\delta \Gamma_{\mu\nu}^\rho = \frac{1}{2} \left( \nabla_\mu {h_{\nu}}^{\rho} + \nabla_\nu {h_{\mu}}^{\rho} - \nabla^{\rho} h_{\mu\nu} \right)
\ee
From this variation and the Palatini equation

\be
\delta {\R^{\rho}}_{\sigma\mu\nu} = \nabla_\mu\ \delta \Gamma_{\nu\sigma}^\rho - \nabla_\nu\ \delta \Gamma_{\mu\sigma}^\rho,  \label{palatini}
\ee
one can easily derive the variation of the Riemann tensor:

\be \label{palatiniexp}
\delta \R_{\rho\sigma\mu\nu} = \frac{1}{2} \left( {\R_{\mu\nu\rho}}^{\lambda} h_{\lambda\sigma} - {\R_{\mu\nu\sigma}}^{\lambda} h_{\lambda \rho} - \nabla_\mu \nabla_\rho h_{\nu\sigma} + \nabla_\mu \nabla_\sigma h_{\nu\rho} - \nabla_\nu \nabla_\sigma h_{\mu\rho} + \nabla_\nu \nabla_\rho h_{\mu\sigma} \right).
\ee

General tensors, of rank at most equal to two, can be uniquely decomposed in tensor, vector and scalar components, according to their tensorial behavior on the $(d-2)$--sphere, the geometry of the black hole event horizon \cite{ik03a}. In particular, this is also true for the perturbations to the metric, but one should note that metric perturbations of tensor type only exist for dimensions $d>4$, unlike perturbations of vector and scalar type, which also exist for $d=4.$ This is because the 2--sphere does not admit any tensor harmonics \cite{Higuchi:1986wu}. Tensor perturbations are therefore intrinsically higher--dimensional.

\subsection{Tensorial perturbations of a spherically symmetric static metric}
\indent

In this work we will only consider tensor type gravitational perturbations to the metric field, for $\a$--corrected $\R^{\mu\nu\rho\sigma} \R_{\mu\nu\rho\sigma}$ black holes in string theory. One should consider perturbations to all the fields present in the low--energy effective action (in our case, the metric and the dilaton), but, as we will show later, one can consistently set tensor type perturbations to the dilaton field to zero. These metric perturbations were studied in \cite{ik03a}, where it is shown that they can be written as

\be \label{htensor}
h_{ij} = 2 r^2 (y^a)\ H_T (y^a)\ \mathcal{T}_{ij} (\theta^i), \quad h_{ia} = 0, \quad h_{ab} = 0,
\ee

\noindent
with $\mathcal{T}_{ij}$ satisfying

\be \label{propt}
\left( \gamma^{kl} D_k D_l + k_T \right) \mathcal{T}_{ij} = 0, \quad D^i \mathcal{T}_{ij} = 0, \quad g^{ij} \mathcal{T}_{ij} = 0.
\ee

\noindent
Here, $D_i$ is the covariant derivative on the $(d-2)$--sphere, associated to the metric $\gamma_{ij}$. Thus, the tensor harmonics $\mathcal{T}_{ij}$ are the eigentensors of the $(d-2)$--laplacian $D^2$, whose eigenvalues are given by $k_T + 2 = \ell \left( \ell + d - 3 \right)$, with $\ell = 2,3,4,\ldots$. It should be further noticed that the expansion coefficient $H_T$ is gauge--invariant by itself. This is rather important: when dealing with linear perturbations to a system with gauge invariance one might always worry that final results could be an artifact of the particular gauge one chooses to work with. Of course the simplest way out of this is to work with gauge--invariant variables, and this is precisely implemented in the Ishibashi--Kodama framework \cite{iks00, ik03a, ik03c}. As it was noticed in \cite{Moura:2006pz}, the Ishibashi--Kodama gauge--invariant variables are also valid for higher derivative theories as long as diffeomorphisms keep implementing gauge transformations. This is because up to now we have only chosen the background metric we wish to perturb: so far, no choice of equations of motion has been done.

Now we consider a static, spherically symmetric background metric. Such a metric is clearly of the type (\ref{ikmetric}), and is given by

\be \label{schwarz}
ds^2 = -f(r)\ dt^2  + g^{-1}(r)\ dr^2 + r^2 d\Omega^2_{d-2}.
\ee

\noindent
The nonzero components of the Riemann tensor for this metric are

\bea \label{ikriemann}
\R_{trtr} &=& \frac{1}{2} f^{\prime \prime} + \frac{1}{4} \frac{f^\prime g^\prime}{g} - \frac{1}{4} \frac{f'^2}{f}\,, \nonumber \\
\R_{itjt} &=& \frac{1}{2} \frac{gf^\prime}{r} g_{ij}\,, \nonumber \\
\R_{irjr} &=& - \frac{1}{2} \frac{g^\prime}{rg} g_{ij}\,, \nonumber \\\
\R_{ijkl} &=& \frac{1}{r^2} \big( 1 - g \big) \Big( g_{ik} g_{jl} - g_{il} g_{jk} \Big).
\eea

\noindent

One first needs to obtain the variation of the Riemann tensor under generic perturbations of the metric. If one collects the expressions for $h_{\mu\nu}$ given in (\ref{htensor}), their covariant derivatives, and further the components of the Riemann tensor given in (\ref{ikriemann}), and replaces them on the Palatini equation (\ref{palatiniexp}), one obtains

\bea
\delta\R_{ijkl} &=& \Big( \big( 3 g - 1 \big) H_T + r g \partial_r H_T \Big) \Big( g_{il} \mathcal{T}_{jk} - g_{ik} \mathcal{T}_{jl} - g_{jl} \mathcal{T}_{ik} + g_{jk} \mathcal{T}_{il} \Big) \nonumber \\
&+& r^2 H_T \Big( D_i D_l \mathcal{T}_{jk} - D_i D_k \mathcal{T}_{jl} - D_j D_l \mathcal{T}_{ik} + D_j D_k \mathcal{T}_{il} \Big), \label{drtensori} \\
\delta\R_{itjt} &=& \left( - r^2 \partial_t^2 H_T + \frac{1}{2} r^2 f f' \partial_r H_T + r f f' H_T \right) \mathcal{T}_{ij}\,, \\
\delta\R_{itjr} &=& \left(- r^2 \partial_t \partial_r H_T -r \partial_t H_T + \frac{1}{2} r^2 \frac{f'}{f} \partial_t H_T \right) \mathcal{T}_{ij}\,, \\
\delta\R_{irjr} &=& \left( - r \frac{g'}{g} H_T - \frac{1}{2} r^2 \frac{g'}{g} \partial_r H_T - 2 r \partial_r H_T - r^2 \partial^2_r H_T \right) \mathcal{T}_{ij}\,, \\
\delta\R_{abcd} &=& 0,
\eea
and further
\bea
\delta \mathcal{R}_{ij} &=& \frac{r^2}{f} \left( \partial^2_t H_T \right) \mathcal{T}_{ij} - r^2 g \left( \partial^2_r H_T \right) \mathcal{T}_{ij} - \frac{1}{2} r^2 \left(f^\prime + g^\prime\right) \left( \partial_r H_T \right) \mathcal{T}_{ij} - r \left(f^\prime + g^\prime\right) H_T \mathcal{T}_{ij} \nonumber \\ &-&\left(d-2\right) r g \left( \partial_r H_T \right) \mathcal{T}_{ij}
+ 2\left(d-3\right) (1-g) H_T \mathcal{T}_{ij} + \left(k_T +2 \right) H_T \mathcal{T}_{ij} \,, \\
\delta \mathcal{R}_{ia} &=& 0, \quad
\delta \mathcal{R}_{ab} = 0, \quad
\delta \mathcal{R} = 0. \label{drtensora}
\eea
\noindent

These are the equations we will need in order to perturb the $\a$--corrected field equations.

\section{Gravitational perturbations to the $\a$--corrected field equations}

\subsection{Analysis on the Einstein frame}
\indent

The $d$--dimensional effective action with $\a$ corrections we will be dealing with is given, in the Einstein frame, by
\be \label{eef} \frac{1}{16 \pi G} \int \sqrt{-g} \left( \R -
\frac{4}{d-2} \left( \d^\mu \phi \right) \d_\mu \phi +
\mbox{e}^{\frac{4}{d-2} \phi} \frac{\lambda}{2}\
\R^{\mu\nu\rho\sigma} \R_{\mu\nu\rho\sigma} \right) \mbox{d}^dx .
\ee

Here $\lambda = \frac{\a}{2}, \frac{\a}{4}$ and $0$, for
bosonic, heterotic and type II strings, respectively. We are only
considering gravitational terms: we can consistently settle all
fermions and gauge fields to zero for the moment. That is not the
case of the dilaton, as it can be seen from the resulting field equations:

\bea
\nabla^2 \phi - \frac{\lambda}{4}\ \mbox{e}^{\frac{4}{2-d} \phi} \left(
\R_{\rho\sigma\lambda\tau} \R^{\rho\sigma\lambda\tau} \right) &=&
0, \label{bdfe} \\ \R_{\mu\nu} + \lambda\ \mbox{e}^{\frac{4}{2-d}
\phi} \left( \R_{\mu\rho\sigma\tau} {\R_{\nu}}^{\rho\sigma\tau} -
\frac{1}{2(d-2)} g_{\mu\nu} \R_{\rho\sigma\lambda\tau}
\R^{\rho\sigma\lambda\tau} \right) &=& 0. \label{bgfe} \eea

The correction term we are considering in (\ref{eef}) is $\R_{\rho\sigma\lambda\tau} \R^{\rho\sigma\lambda\tau}$, the square of the Riemann tensor, which we generically designate as $\R^2$: since we are not considering the Ricci tensor in the corrections (it would only contribute at a higher order in $\lambda$), there is no possible confusion. From (\ref{bdfe}) one sees that this correction term acts as a source for the dilaton and, therefore, one cannot set the dilaton to zero without setting this term to zero too. Still, as it was shown in \cite{Moura:2009it} and we will review later (see eq. (\ref{fr2})), for a spherically symmetric metric like (\ref{schwarz}), at order $\lambda=0$ the dilaton is a constant (which can be always set to 0). The dilaton only gets nonconstant terms at order $\lambda;$ this is why we could neglect terms which are quadratic in $\phi$ while deriving these field equations, since we are only working perturbatively to first order in $\lambda.$

In the present context, any black hole solution is built perturbatively in $\lambda,$ and a solution will only be valid in regions where $r^2 \gg \lambda$, \textit{i.e.}, any perturbative solution is only valid for black holes whose event horizon is much bigger than the string length.

We want to study scattering processes associated to solutions to the field equations above and, therefore, they are the ones which we will perturb.

The perturbation of the $\a$--corrected field equations (\ref{bdfe}) and (\ref{bgfe}) has already been taken in \cite{Moura:2006pz}, for a spherically symmetric metric like (\ref{schwarz}) but with $f(r)=g(r);$ here we consider the general case (\ref{schwarz}) and apply the results to concrete metrics.

By perturbing (\ref{bdfe}) and (\ref{bgfe}) one gets

\bea
\delta \nabla^2 \phi &-& \frac{\lambda}{4}\ \mbox{e}^{\frac{4}{2-d} \phi}\ \delta \left( \R_{\rho\sigma\lambda\tau} \R^{\rho\sigma\lambda\tau} \right) + \frac{\lambda}{d-2}\ \mbox{e}^{\frac{4}{2-d} \phi}\ \R_{\rho\sigma\lambda\tau} \R^{\rho\sigma\lambda\tau}\ \delta \phi = 0, \label{pbdfe} \\
\delta \R_{ij} &+& \lambda\ \mbox{e}^{\frac{4}{2-d} \phi} \left[ \delta \left( \R_{i\rho\sigma\tau} {\R_{j}}^{\rho\sigma\tau} \right) - \frac{1}{2(d-2)} \R_{\rho\sigma\lambda\tau} \R^{\rho\sigma\lambda\tau}\ h_{ij} \right. \nonumber \\
&-& \left. \frac{1}{2(d-2)}\ g_{ij}\ \delta \left( \R_{\rho\sigma\lambda\tau} \R^{\rho\sigma\lambda\tau} \right) \right] + \frac{4}{d-2}\ \R_{ij}\ \delta \phi = 0. \label{pbgfe}
\eea

\noindent
Using the explicit form of the Riemann tensor (\ref{ikriemann}) together with the variations (\ref{htensor}) and (\ref{drtensori}--\ref{drtensora}), one can compute the terms in (\ref{pbdfe}) and (\ref{pbgfe}).

Using these variations, it is a simple computation to verify that $\delta \left( \R_{\rho\sigma\lambda\tau} \R^{\rho\sigma\lambda\tau} \right) \equiv 0.$ From this fact and (\ref{pbdfe}), we see that one can consistently set $\delta \phi=0,$ as expected for a tensorial perturbation of a scalar field. The derivation is explicitly given in article \cite{Moura:2006pz}, to which we refer the reader. Eq. (\ref{pbdfe}) does not give us any other relevant information.

Collecting the several expressions, the result for (\ref{pbgfe}) finally becomes

\bea
&&
\left( 1 - 2 \lambda \frac{f'}{r} \right) \frac{r^2}{f} \partial^2_t H_T - \left( 1 - 2 \lambda \frac{g'}{r} \right) r^2 g \,\partial^2_r H_T \nonumber \\
&&
- \left[(d-2) r g + \frac{1}{2} r^2 \left(f'+g'\right) + 4 \lambda (d-4) \frac{g \left( 1 - g \right)}{r} - 4 \lambda g g' - \lambda r \left( f'^2 + g'^2 \right) \right] \partial_r H_T \nonumber \\
&&
+ \left[ \left( \ell \left( \ell + d - 3 \right) -2 \right) \left(1+\frac{4\lambda}{r^2}\left(1-g\right)\right) + 2(d-2)- 2(d-3)g- r\left(f' + g'\right) \right. \nonumber \\
&&
+ \left. \lambda \left( 8 \frac{1 - g}{r^2} + 2 \left( d - 3 \right)\frac{\left( 1 - g \right)^2}{r^2} - \frac{r^2}{d-2} \left[f'' + \frac{1}{2} \left( \frac{f' g'}{g} - \frac{f'^2}{f}\right)\right]^2 \right) \right] H_T = 0. \label{master0}
\eea

\noindent
This is a second order partial differential equation for the perturbation function $H_T.$ If we now divide (\ref{master0}) by $\left( 1 - 2 \lambda \frac{f'}{r} \right) \frac{r^2}{f},$ we obtain an equation of the form
\be \label{dottigen}
\partial^2_t H_T - F^2(r)\ \partial^2_r H_T + P(r)\ \partial_r H_T + Q(r)\ H_T = 0
\ee
with
\bea
F &=& \sqrt{\frac{1 - 2 \lambda \frac{g'}{r}}{1 - 2 \lambda \frac{f'}{r}} fg}, \nonumber \\
P &=& - \frac{f}{1 - 2 \lambda \frac{f'}{r}} \left[ (d-2) \frac{g}{r} + \frac{1}{2} \left(f'+g'\right) + \frac{2 \lambda}{r} \left( 2 ( d - 4 ) \frac{g \left( 1 - g \right)}{r^2}
- 2 \frac{g g'}{r} - \frac{1}{2} \left(f'^2+g'^2\right) \right) \right], \nonumber \\
Q &=& \frac{f}{1 - 2 \lambda \frac{f'}{r}} \left[ \frac{\ell \left( \ell + d - 3 \right)}{r^2} - \frac{f'+g'}{r} + 2 (d-3) \frac{1-g}{r^2} \right. \nonumber \\ &+& \left. \frac{\lambda}{r^2} \left( 4 \ell \left( \ell + d - 3 \right) \frac{1-g}{r^2} + 2 (d-3) \frac{\left( 1-g \right)^2}{r^2} - \frac{r^2}{d-2} \left[f'' + \frac{1}{2} \left( \frac{f' g'}{g} - \frac{f'^2}{f}\right)\right]^2
\right) \right]. \label{fpqnp}
\eea
\noindent
For our purposes, we would like to re--write the above equation (\ref{dottigen}) in a more tractable form, as a master equation. In order to achieve so, we follow a procedure similar to the one in \cite{dg05a}, defining a gauge--invariant ``master variable'' for the gravitational perturbation as

\be
\Phi = k(r) H_T, \quad k(r) = \frac{1}{\sqrt{F}} \exp \left( - \int \, \frac{P}{2F^2} \, dr \right), \label{k}
\ee
and replacing $\partial/\partial r$ by $\partial/\partial r_*,$ $r_*$ being the tortoise coordinate defined in this case by $dr_* = \frac{dr}{F(r)}.$ It is then easy to see that an equation like (\ref{dottigen}) may be written as a master equation:

\be
\frac{\partial^2 \Phi}{\partial r_*^2} - \frac{\partial^2 \Phi }{\partial t^2} = \left( Q + \frac{F'^2}{4} - \frac{F F''}{2} - \frac{P'}{2} + \frac{P^2}{4 F^2} + \frac{P F'}{F} \right) \Phi \equiv V_{\textsf{T}} \left[ f(r), g(r) \right] \Phi. \label{potential0}
\ee
This is the equation we shall be working with for the rest of the article. We see that the field equations for the tensorial perturbations of a background spherically symmetric metric like (\ref{schwarz}) in $d$ dimensions can be reduced to a single second order partial differential equation, in $(r_*,t)$ coordinates, for a master variable $\Phi$ which is a simple combination of gauge--invariant variables. This result had been obtained in \cite{ik03a} for Einstein gravity; here, we see that it is also valid in the presence of string corrections quadratic in the Riemann tensor.

In order to explicitly compute the potential $V_{\textsf{T}} \left[ f(r), g(r) \right],$ one should first simplify the expressions from (\ref{fpqnp}). To begin with, one can judiciously use the field equation (\ref{bgfe}) to derive the relation
$$\lambda \R_{abcd} \R^{abcd} = 2 g^{ij} \R_{ij} + \lambda \R_{ijkl} \R^{ijkl}.$$
Here we used the fact that, in the presence of a metric like (\ref{schwarz}), the dilaton field is of order $\lambda,$ as it was explicitly shown in \cite{Moura:2009it}. This way we could neglect all the dilaton terms, which would only contribute at least to order ${\mathcal{O}} (\lambda^2)$). Using the above relation and the explicit form of the Riemann tensor (\ref{ikriemann}), we can obtain the (on--shell) relation
\be \label{f''}
\lambda \left[f'' + \frac{1}{2} \left( \frac{f' g'}{g} - \frac{f'^2}{f}\right)\right]^2 -\frac{2 (d-3) (d-2)}{r^2} \left(\frac{\lambda (1-g)}{r^2}+1\right) (1-g)+\frac{(d-2)}{r} \left(\frac{g f'}{f}+g'\right)=0.
\ee
We may use the relation above in order to remove the $\left[f'' + \frac{1}{2} \left( \frac{f' g'}{g} - \frac{f'^2}{f}\right)\right]^2$ term from $Q$ in (\ref{fpqnp}). Also, although the expressions in (\ref{fpqnp}) are non--polynomial in $\lambda$, we can expand them and take only the first order terms, since that is the order in $\lambda$ to which we are working. A simple power series expansion yields

\bea
F &=& \sqrt{fg}\left(1 +\lambda \frac{f'-g'}{r}\right), \nonumber \\
P &=& - f \left[ (d-2) \frac{g}{r} + \frac{1}{2} \left(f'+g'\right) + \frac{\lambda}{r^2}  \left(4 (d-4) \frac{g(1-g)}{r}+r g'\left(f'-g'\right)-4 g g' +2 (d-2) g f' \right) \right], \nonumber \\
Q &=& \frac{\ell \left( \ell + d - 3 \right)}{r^2} f + \frac{(g-f)f'}{r} +
\frac{2\lambda}{r^2} \left[ \frac{\ell \left( \ell + d - 3 \right)}{r} f \left( 2 \frac{1-g}{r} + f' \right) + (g-f) f'^2 \right].
\label{fpq}
\eea
Replacing the expressions from (\ref{fpq}) on $V_{\textsf{T}} \left[ f(r), g(r) \right]$ given in (\ref{potential0}), one finally obtains

\bea V_{\textsf{T}} [f(r),g(r)] &=&
\frac{1}{16 r^2 f g}
\left[(16 \ell (\ell +d-3) f^2 g+ r^2 f^2 f'^2 +3 r^2 g^2 f'^2-2 r^2 f (f+g) f' g' -4 r^2 fg (g-f) f'' \right. \nonumber \\
&+& \left. 16r f g^2 f' +4 r (d-6)f^2 g f' +4 (d-2)r f^2 g g' +4 (d-4) (d-2) f^2 g^2 \right] \nonumber \\
&+&\frac{\lambda}{8 r^4 f g} \left[32 \ell (\ell +d-3) f^2 (1-g) g +16 \ell (d+\ell -3) f^2 g f' r -r^3 f^2 f'^2 \left(f'-g'\right) \right. \nonumber \\
&+& 3 r^3 g^2 f'^2 \left(f'-g'\right)-2 r^3 f g f' \left(f'-g'\right) g' -4 r^3 f^2 g f' \left(f''-g''\right)-2 r^3 f^2 g g' \left(f''-g''\right) \nonumber \\
&+& 2 r^3 f g^2 \left(-3 f' f''+2 g' f''+f' g''\right) -4 r^3 f^2 g^2 \left(f'''-g'''\right) +18 r^2 f g^2 f'^2 -12 r^2 f^2 g f'^2 \nonumber \\
&-& 10 r^2 f^2 g g'^2 -2 r^2 f g^2 f' g' +2 r^2 (4 d-13) f^2 g f' g' +8 r^2 f^2 g^2 f'' +8 (d-5) r^2 f^2 g^2 g''
 \nonumber \\
&+& 4 r(d-4)^2 f^2 g^2 (f' + g') +8r f^2 g^2(g'-f') +8 (d-4) r f^2 g (f' + g'-4g g') \nonumber \\
&+& \left. 16 (d-5) (d-4) f^2 g^2 (1-g) \right].
\label{potential}
\eea
Taking $f=g$ in (\ref{schwarz}), (\ref{potential}) matches the result of \cite{Moura:2006pz}, as it should. Also, at order $\lambda=0,$ (\ref{potential}) matches the Einstein-Hilbert potential obtained in \cite{ik03a}.

Equation (\ref{potential}) gives the generic expression for the potential for tensor--type gravitational perturbations of any kind of static, spherically symmetric $\R^2$ string--corrected black hole in $d$--dimensions of the form (\ref{schwarz}). This is also one of the main results of this article. Knowing this potential, we are now ready to study the stability of such black holes under those perturbations.

\subsection{Analysis in different frames}
\label{diffr}
\indent

Under a conformal transformation, the metric and Riemann tensor transform as:
\bea \label{confgen}
g_{\mu\nu} &\rightarrow& \exp \left( \Omega \right) g_{\mu\nu}, \\
{\R_{\mu\nu}}^{\rho\sigma} &\rightarrow& \exp \left( -\Omega \right) \left( {\R_{\mu\nu}}^{\rho\sigma} - 2 {\delta_{\left[\mu\right.}}^{\left[\rho\right.} \nabla_{\left.\nu \right]} \nabla^{\left.\sigma \right]} \Omega + {\delta_{\left[\mu\right.}}^{\left[\rho\right.} \left(\nabla_{\left.\nu \right]} \Omega \right) \nabla^{\left.\sigma \right]} \Omega - {\delta_{\left[\mu\right.}}^{\left[\rho\right.} \delta_{\left.\nu\right]}^{\left. \, \, \, \, \sigma\right]} \left(\nabla^{\lambda}\Omega\right) \nabla_{\lambda}\Omega \right).
\eea

In our perturbation analysis we assumed an action/solution taken in the Einstein frame. But the action obtained directly from string perturbation theory comes in the string frame as
\be \label{esf}
\frac{1}{16 \pi G} \int \sqrt{-g}\ \mbox{e}^{-2 \phi} \Big( \R + 4 \left( \d^\mu \phi \right) \d_\mu \phi + \frac{\lambda}{2}\ \R^{\mu\nu\rho\sigma} \R_{\mu\nu\rho\sigma} \Big) \mbox{d}^dx.
\ee
Taking a dilaton--dependent conformal transformation (\ref{confgen}) with $\Omega=\frac{4}{d-2} \phi,$ we obtain from (\ref{esf}) the action (\ref{eef}) in the Einstein frame, plus some terms with derivatives of the dilaton, which would only contribute at higher orders in $\lambda,$ since as we saw the dilaton itself is already of order $\lambda.$

In general, if $Y(\R)$ is a scalar polynomial in the Riemann tensor representing the higher derivative corrections, with conformal weight $w$ and the convention that $w \left( g_{\mu\nu} \right) = +1,$ an arbitrary action with $\a$ corrections is (just the gravitational sector)
\be
\frac{1}{16 \pi G} \int \sqrt{-g} \Big( \R - \frac{4}{d-2} \left( \d^\mu \phi \right) \d_\mu \phi + z Y(\R) \Big) \mbox{d}^dx,
\ee
with $z$ being, up to a numerical factor, the suitable power of the inverse string tension $\a$ for $Y(\R)$.

In a different frame, after the conformal transformation (\ref{confgen}), this action becomes
\be
\frac{1}{16 \pi G} \int \sqrt{-g}\ \mbox{e}^{\frac{d-2}{2} \Omega} \left( \widetilde{\R} - \frac{4}{d-2} \left( \d^\mu \phi \right) \d_\mu \phi + z\ \mbox{e}^{\left( 1 + w \right) \Omega} Y(\widetilde{\R}) \right) \mbox{d}^dx.
\ee
Again, if the conformal transformation (\ref{confgen}) is dilaton--dependent, it will generate from $Y(\widetilde{\R})$ some dilaton terms which are of higher order in $\lambda.$ Besides that, other lower order parcels containing the dilaton may be generated, affecting the numerical factor (and sign) of its kinetic term, as one can see comparing (\ref{esf}) to (\ref{eef}). These dilaton terms are the only change in the graviton field equation generated by a dilaton--dependent conformal transformation (\ref{confgen}).

The field equation (\ref{bgfe}) we considered was obtained after removing the trace term $-\frac{1}{2} g_{\mu\nu} \R,$ obtained after contracting (\ref{bgfe}) itself. This procedure is totally independent of the chosen frame. Other dilaton terms were removed by considering the dilaton field equation, which obviously depends on the choice of frame (equation (\ref{bdfe}) is valid in the Einstein frame). Still, as we have seen under tensorial perturbations the dilaton $\phi$ is inert, and so is its field equation, in any chosen frame. Indeed the arguments in the discussion after (\ref{bgfe}), namely the fact that $\delta \left( \R_{\rho\sigma\lambda\tau} \R^{\rho\sigma\lambda\tau} \right) \equiv 0$, do not depend on the choice of frame. This means that, in order to study tensorial perturbations, only (\ref{bgfe}) is relevant, and the results obtained from it are valid in any chosen frame, namely the equation governing tensorial perturbations of the metric (\ref{master0}) (or, equivalently, (\ref{potential0})).

Finally we should mention that, although the dilaton field equation looks different in the string and in the Einstein frames (see \cite{cmp89}), their spherically symmetric $d$--dimensional solution is the same in both frames. Indeed the string frame dilaton solution obtained in \cite{cmp89} is equivalent to the Einstein frame dilaton solution from \cite{Moura:2009it}, apart from the normalization at infinity (which in \cite{Moura:2009it} was taken to be zero - this is the solution we use). This is not obvious if one looks at the two solutions, but it can be verified case by case in $d$ using symbolic computation software. The reason becomes clear after comparing the two field equations and neglecting terms of higher order in $\lambda.$ This fact makes the change from the string to the Einstein frame (and vice versa) unambiguous.

\section{General analysis of perturbative stability}
\indent

In order to study the stability of a solution, we use the ``S--deformation approach'', first introduced in \cite{ik03b} and later further developed in \cite{dg04, dg05a}. Let us briefly review this technique in the following (for more details we refer the reader to the original discussion in \cite{ik03b}).

After having obtained the potential $V_{\textsf{T}}$ for the master equation (\ref{potential0}), one assumes that its solutions are of the form $\Phi(x,t) = e^{i\omega t} \phi(x)$, such that $\frac{\partial\Phi}{\partial t} = i\omega \Phi$. In this way the master equation (\ref{potential0}) may be written in Schr\"odinger form, for a generic potential $V(x),$ as

\be
\left[ - \frac{d^2}{dx^2} + V(x) \right] \phi(x) \equiv A \phi(x) = \omega^2
\phi(x).
\ee

\noindent
A given solution of the gravitational field equations will then be perturbatively stable if and only if the operator $A$ defined above has no negative eigenvalues for $x \in {\mathbb{R}}$ \cite{ik03b}. The above condition is equivalent to the positivity, for any given $\phi,$ of the inner product \cite{ik03b}

\be
\left \langle \phi \left| A \phi \right \rangle \right. = \int_{-\infty}^{+\infty} \phi^\dagger (x) \left[ - \frac{d^2}{dx^2} + V(x) \right] \phi(x)\ dx,
\ee

\noindent
which, after some integrations by parts and further algebra, may be rewritten as

\be
\left \langle \phi \left| A \phi \right \rangle \right.  = \int_{-\infty}^{+\infty} \left[ \left| \frac{d\phi}{dx} \right|^2 + V(x) \left|\phi \right|^2 \right] dx = \int_{-\infty}^{+\infty} \left[ \left| D \phi \right|^2 + \widetilde{V}(x) \left| \phi \right|^2 \right] dx.
\ee

\noindent
Here we have defined $D = \frac{d}{dx} + S$ and $\widetilde{V}(x) = V(x) + f \frac{dS}{dr} - S^2$, with $S$ a completely arbitrary function. Taking $S = - \frac{F}{k} \frac{dk}{dr},$ with $k(r)$ given by (\ref{k}), we simply obtain $\widetilde{V}(x) = Q$ and

\be
\left \langle \phi \left| A \phi \right \rangle \right. = \int_{-\infty}^{+\infty} \left|D \phi \right|^2 dx + \int_{-\infty}^{+\infty} Q(x) \left|\phi \right|^2 dx.
\ee

\noindent
The second term of the expression above may be written as ($R_H$ being the radius of the event horizon)

\be
\int_{R_H}^{+\infty} \frac{Q(r)}{F(r)} \left|\phi \right|^2 dr,
\ee

\noindent
Since $\left|\phi \right|^2$ and $\left|D \phi \right|^2$ are positive, perturbative stability of a given black hole solution then follows if one can prove that $\frac{Q(r)}{F(r)}$ is a positive function for $r \ge R_H.$ Here we note that, by definition, the horizon is the largest root of $f(r),$ i.e $f>0$ for $r>R_H.$ Here we assume the same is valid for $g,$ i.e. the classical part of $F=F_0 + \lambda F_1,$ given by $F_0=\sqrt{fg},$ is well defined and positive for $r>R_H.$ This does not mean that $F$ is necessarily positive: close to the horizon, $F_0=\sqrt{fg}$ is expected to have very small values, which we cannot assume to be larger than the $\lambda$--corrections, as one usually does. These corrections, from (\ref{fpq}) given by $F_1= \sqrt{fg} \frac{f'-g'}{r},$ may well be negative if $f'<g',$ i.e. if $f$ grows slower than $g.$ This means that, very close to the horizon, for a small range, it may happen that $F$ is negative (that would not affect the metric, as $F$ is not part of it). This way, it is more rigorous to verify the sign of $\frac{Q(r)}{F(r)}.$ From (\ref{fpq}) we have, to first order in $\lambda,$
\bea
\frac{Q(r)}{F(r)} &=& \frac{1}{r^2 \sqrt{fg}} \Big[ \ell \left( \ell + d - 3 \right)f + r (g-f)f' \nonumber \\
&+& \frac{\lambda}{r^2} \left[ \ell \left( \ell + d - 3 \right) f \left( 4 (1-g) +r (f'+g') \right) + r^2 (g-f) f' (f'+g') \right] \Big].
\label{qf}
\eea

\noindent
For a given black hole solution of (\ref{bgfe}), the positiveness of this expression will assure its stability under tensor perturbations of the metric.

\section{Application to concrete string-corrected black hole solutions}
\indent

We now apply our result to the study of the stability of a few specific black hole solutions in string theory. Articles \cite{dg04,dg05a} were the first to apply this formalism to black holes with $\R^2$ corrections in $d$ dimensions, but to solutions which are not in the context of string theory.

The most general static, spherically symmetric metric in $d$ dimensions for pure Einstein-Hilbert
gravity in vacuum is the Tangherlini solution \cite{Tangherlini:1963bw}, of the form (\ref{schwarz}) with $f(r)=g(r) \equiv f_0^T(r),$ where
\be
f_0^T(r) =1 - \left(\frac{R_H}{r}\right)^{d-3}, \label{tangher}
\ee
$R_H$ being the horizon radius.

The first solution of such kind with string theoretical $\R^2$ corrections was obtained by Callan, Myers and Perry \cite{cmp89}. It is of the form (\ref{schwarz}), with $f(r)=g(r) \equiv g_{CMP}(r),$ where
\be
g_{CMP}(r) = \left(1-\left(\frac{R_H}{r}\right)^{d-3} \right) \left[1- \lambda
\frac{(d-3)(d-4)}{2} \frac{R_H^{d-5}}{r^{d-1}}
\frac{r^{d-1}-R_H^{d-1}}{r^{d-3} - R_H^{d-3}} \right]. \label{fr2x}
\ee
It is clearly a deformation of the Tangherlini metric (\ref{tangher}). The stability of such solution under tensor perturbations was proven in \cite{Moura:2006pz}.

Next we will apply the method to two previously unstudied cases.

\subsection{The string-corrected dilatonic $d$--dimensional black hole}
\subsubsection{Description of the solution}
\indent

The Callan--Myers--Perry solution expresses the effect of the string $\R^2$ corrections, but it does not consider any other string effects, namely the fact that string theories live in $d_S$ spacetime dimensons ($d_S=10$ or 26 on heterotic or bosonic strings, respectively), and have to be compactified to $d$ dimensions on a $d_S - d$--dimensional manifold. When passing from the string to the Einstein frame, the volume of the compactification manifold becomes spatially varying. In the simple case when such manifold is a flat torus, that volume depends only on the $d$--dimensional part of the dilaton $\phi$ and, after solving the $\a$--corrected field equation (\ref{pbgfe}) the metrics of the compactification manifold and of the $d$--dimensional spacetime decouple.

The explicit solution was worked out in \cite{Moura:2009it}. The $d$--dimensional part of the metric is of the form (\ref{schwarz}), with $g(r)$ being equal to $g_{CMP}(r)$ given by (\ref{fr2x}) and
\be
f(r) = g(r) + 4 \left( 1 -
\left(\frac{R_H}{r}\right)^{d-3} \right)
\frac{d_s-d}{\left(d_s-2\right)^2} \left(\phi - r \phi'\right).
\label{fnew}
\ee

The general solution for the dilaton, in the classical background of the spherically symmetric Tangherlini black hole (\ref{tangher}), was also obtained in \cite{Moura:2009it}. As we previously mentioned, this solution is necessarily of order $\lambda:$ it is given by $\phi(r) = \frac{\lambda}{R_H^2} \varphi(r),$ with
\bea \label{fr2}
\varphi(r) &:=& \frac{(d-2)^2}{4}
\left[\ln\left( 1 - \left(\frac{R_H}{r}\right)^{d-3} \right) -\frac{d-3}{2}\left(\frac{R_H}{r}\right)^2 - \frac{d-3}{d-1}
\left(\frac{R_H}{r}\right)^{d-1} \right. \nonumber \\
&+& \left. B \left(\left(\frac{R_H}{r}\right)^{d-3};\, \frac{2}{d-3}, 0 \right)\right]
\eea
with $B(x;\,a,b)=\int_0^x t^{a-1}\,(1-t)^{b-1}\,dt \!$ being the incomplete Euler beta function. Its derivative is given by
\be
\varphi'\left(r\right)=\frac{(d-3)
(d-2)^2}{4} \frac{R_H^{d-3}}{r^{d-2}}
\frac{1-\left(\frac{R_H}{r}\right)^{d-1}}{1-\left(\frac{R_H}{r}\right)^{d-3}}. \label{fr3}
\ee

Both $\varphi(r)$ and $\varphi'(r)$ are well--defined functions everywhere, namely at the horizon: \footnote{The digamma function is given by $\psi(z)=\Gamma'(z)/\Gamma(z),$ $\Gamma(z)$ being the usual $\Gamma$ function. For positive $n,$ one defines $\psi^{(n)}(z)=d^n\,\psi(z)/d\,z^n.$ This definition can be extended for other values of $n$ by fractional calculus analytic continuation. These are meromorphic functions of $z$ with no branch cut discontinuities.

$\gamma$ is Euler's constant, defined by $\gamma=\lim_{n \to \infty} \left(\sum_{k=1}^n \frac{1}{k} - \ln n \right),$ with numerical value $\gamma \approx 0.577216.$}
\bea
\varphi\left(r_H\right)&=&- \frac{(d-2)^2}{8 (d-1)} \left(d^2-2d+2 (d-1) \left(\psi^{(0)}\left(\frac{2}{d-3}\right) + \gamma \right) -3\right), \label{firh} \\
\varphi'\left(r_H\right)&=& \frac{(d-2)^2 (d-1)}{4 R_H} \label{firhd}.
\eea

The derivative (\ref{fr3}) is a positive function, which means the dilaton solution (\ref{fr2}) is negative and strictly monotonic. It grows up to 0 at infinity, as it can be seen from the asymptotic expansion
\be
\varphi(r) \approx -\frac{(d-2)^2}{4} \left(\frac{R_H}{r}\right)^{d-3}
-\frac{(d-2)^2}{8} \left(\frac{R_H}{r}\right)^{2d-6}
+ \frac{(d-2)(d-3)}{8} \left(\frac{R_H}{r}\right)^{2d-4}+ {\mathcal{O}} \left( \left( \frac{R_H}{r}\right)^{2d-3} \right),
\label{dildeser}
\ee
obtained from
\bea
&& B \left(\left(\frac{R_H}{r}\right)^{d-3};\, \frac{2}{d-3}, 0 \right)+ \ln\left( 1 - \left(\frac{R_H}{r}\right)^{d-3} \right) \nonumber \\
&=& \frac{d-3}{2} \left(\frac{R_H}{r}\right)^2 - \left(\frac{R_H}{r}\right)^{d-3} + \frac{d-3}{d-1} \left(\frac{R_H}{r}\right)^{d-1}-\frac{1}{2} \left(\frac{R_H}{r}\right)^{2(d-3)} + {\mathcal{O}} \left( \left( \frac{R_H}{r}\right)^{2d-5} \right).
\eea

\subsubsection{Study of the stability}
\indent

In order to check for the stability of this metric under tensor perturbations, we split $\frac{Q}{F},$ given in (\ref{qf}), in their "classical" and $\lambda$--corrected parts: $\frac{Q}{F} = \left.\frac{Q}{F}\right|_0 + \lambda \left.\frac{Q}{F}\right|_1.$ Since $\left.\frac{Q}{F}\right|_1$ is already multiplied by $\lambda,$ the order at which we are working, it must be evaluated using the functions corresponding to the $\lambda=0$ Tangherlini solution (\ref{tangher}). Using $f(r)=g(r) = f_0^T(r),$ and since for $r>R_H$ clearly $f_0^T >0,$ from (\ref{qf}) we get
$$r^4 \sqrt{fg} \left.\frac{Q}{F}\right|_1 = \frac{2}{r^2} \frac{\ell \left( \ell + d - 3 \right)}{r} f_0^T \left( 2 \frac{1-f_0^T}{r} + f_0^{T'} \right).$$
We only need to compute
$$2 \frac{1-f_0^T}{r} + f_0^{T'}= (d-3) \frac{R_H^{d-3}}{r^{d-2}},$$
which is always a positive quantity. This way, $\left.\frac{Q}{F}\right|_1$ is always positive.

$r^2 \sqrt{fg} \left.\frac{Q}{F}\right|_0=\frac{\ell \left( \ell + d - 3 \right)}{r^2} f + \frac{(g-f)f'}{r} $ must be computed with the full, $\lambda$--corrected metric (therefore, it will also include $\lambda$ corrections).

By definition, the horizon is the largest root of $f(r).$ Since asymptotically $f(r) \rightarrow 1,$ $f(r)$ must be positive for $r>R_H.$

As we have seen, the dilaton solution (\ref{fr2}) is negative and its derivative (\ref{fr3}) is a positive function.
Therefore $\phi - r \phi'$ is a negative quantity. Since $1 - \left(\frac{R_H}{r}\right)^{d-3} >0,$ from (\ref{fnew}) we get $g-f>0$ (always for $r>R_H$). Also from (\ref{fnew}) $g-f$ is of order $\lambda,$ i.e. its classical part is zero. This way, in the term $(g-f)f'$ of $\left.\frac{Q}{F}\right|_0,$ one can take just the classical part $f_0^{T'}$ to compute $f'$: its $\lambda$ correction would only contribute to order $\lambda^2.$ But, as we know, $f_0^{T'}>0.$ This means $\left.\frac{Q}{F}\right|_0$ (and therefore $\frac{Q}{F}$) is indeed positive for $r>R_H.$ This way, the dilatonic $d$--dimensional black hole solution (\ref{fnew}) is indeed stable under tensorial perturbations.

\subsection{The double--charged black hole}
\label{giveon}
\subsubsection{Description of the solution}
\indent

In article \cite{Giveon:2009da} one can find black holes in any dimension formed by a fundamental string compactified on an internal circle with any momentum $n$ and winding $w,$ both at leading order and with leading $\a$
corrections. One starts with the Callan--Myers--Perry solution in the string frame \cite{cmp89}, which is of the form (\ref{schwarz}), with $f, g$ replaced by $f^{CMP}_S, g^{CMP}_S,$ given by
\bea
f^{CMP}_S(r)&=&f_0^T \left(1+2 \frac{\lambda}{R_H^2} \mu(r)\right), \nonumber \\
g^{CMP}_S(r)&=&f_0^T \left( 1-2 \frac{\lambda}{R_H^2} \epsilon(r)\right), \nonumber \\
\epsilon(r)&=&\frac{d-3}{4} \frac{\left(\frac{R_H}{r}\right)^{d-3}}{1-\left(\frac{R_H}{r}\right)^{d-3}} \left[\frac{(d-2)(d-3)}{2}
-\frac{2(2d-3)}{d-1} + (d-2)\left(\psi^{(0)}\left(\frac{2}{d-3}\right) + \gamma \right)\right. \nonumber \\ &+& \left. d \left(\frac{R_H}{r}\right)^{d-1} +\frac{4}{d-2} \varphi(r) \right], \label{ep} \\
\mu(r)&=&-\epsilon(r)+\frac{2}{d-2} (\varphi(r)-r \varphi'(r)). \label{mu}
\eea
$f_0^T$ is given by (\ref{tangher}) and $\varphi(r)$ is given by (\ref{fr2}). This dilaton solution $\varphi(r)$ is such that $\epsilon(r), \mu(r)$ are finite at the horizon $r=R_H$ \cite{Moura:2009it}: indeed, from (\ref{firh}), (\ref{firhd}) and the definitions (\ref{ep}), (\ref{mu}) one gets
\be
\epsilon(R_H)= -\frac{d-1}{2}, \, \mu(R_H)=-\frac{3d\left(d-3\right)\left(d-\frac{5}{3}\right)}{4\left(d-1\right)}
-\frac{d-2}{2}\left(\psi^{(0)}\left(\frac{2}{d-3}\right) + \gamma \right).
\ee

In the same way as $\varphi(r)$, also $\epsilon(r)$ in (\ref{ep}) is negative (always for $r>R_H$): it has a negative value at the horizon and grows up to 0 at asymptotic infinity. Indeed one can easily show (for instance, constructing a table with Mathematica) that, for relevant values of $d,$ the first line in (\ref{ep}) obeys $\frac{(d-2)(d-3)}{2}-\frac{2(2d-3)}{d-1} + (d-2)\left(\psi^{(0)}\left(\frac{2}{d-3}\right) + \gamma \right)<0.$ There is a positive contribution from $d \left(\frac{R_H}{r}\right)^{d-1}$, but it decreases with increasing $r,$ differently from other negative contributions. Another remarkable fact from $\epsilon(r)$ is that, for most of its range, it has very small numerical values (when compared to $\varphi(r)$). This is due to the overall factor $\frac{\left(\frac{R_H}{r}\right)^{d-3}}{1-\left(\frac{R_H}{r}\right)^{d-3}},$ because of which the absolute value of $\epsilon(r)$ decreases much faster than that of $\varphi(r).$ Because of that, $\mu(r)$ which, from (\ref{mu}), is a difference of negative quantities, is also negative in all its range. This negativeness of $\epsilon(r), \mu(r)$ can be checked simply by plotting these two functions of $r/R_H$ for all relevant values of $d.$

This Callan--Myers--Perry metric in the string frame is lifted to an additional dimension by adding an extra coordinate, taken to be compact (this means to produce a uniform black string). One then performs a boost along this extra direction, with parameter $\alpha_w$, and $T$--dualizes around it (to change string momentum into winding), obtaining a $(d+1)$--dimensional black string winding around a circle. Finally one boosts one other time along this extra direction, with parameter $\alpha_p$, in order to add back momentum charge. One finally obtains a spherically symmetric black hole in $d$ dimensions with two electrical charges.

The whole process is worked out in detail in \cite{Giveon:2009da}; the final metric, in the string frame, is of the form (\ref{schwarz}), with $f, g$ given by
\bea
f_S(r)&=&\frac{f_0^T}{\Delta(\alpha_n)\Delta(\alpha_w)}
\left[1+\frac{2 \lambda}{R_H^2} \frac{\mu(r)}{\Delta(\alpha_n)\Delta(\alpha_w)}
-\frac{2 \lambda}{R_H^2} \mu(r) \frac{\sinh^2(\alpha_n)\sinh^2(\alpha_w)}{\Delta(\alpha_n)\Delta(\alpha_w)} \left(\frac{R_H}{r}\right)^{2(d-3)}\right. \\
&+&\frac{2 \lambda}{R_H^2} \mu(r)\left(\frac{\sinh^{2}\alpha_n}{\Delta(\alpha_n)}
+\frac{\sinh^{2}\alpha_w}{\Delta(\alpha_w)}\right)+\left. \frac{\lambda}{R_H^2} (d-3)^2 f_0^T \left(\frac{R_H}{r}\right)^{2(d-2)}
\frac{\sinh^2(\alpha_n)\sinh^2(\alpha_w)}{\Delta(\alpha_n)\Delta(\alpha_w)}\right], \nonumber\\
\Delta\left(x\right)&:=&1+\left(\frac{R_H}{r}\right)^{d-3}\sinh^2x, \label{delta} \\
g_S(r)&=&f_0^T\left( 1-2\frac{\lambda}{R_H^2} \epsilon(r)\right).
\eea
The dilaton in this case is given by

\bea
e^{-2\phi}&=&\sqrt{\Delta(\alpha_n)\Delta(\alpha_w)}
\left[1-2 \frac{\lambda}{R_H^2} \varphi(r)-\frac{\lambda}{R_H^2} \mu(r) f_0^T \left(\frac{\sinh^2\alpha_n}{\Delta(\alpha_n)}
+\frac{\sinh^2\alpha_w}{\Delta(\alpha_w)}\right) \right. \nonumber\\
&-&\left.\frac{\lambda}{R_H^2} \frac{(d-3)^2}{2} f_0^T \left(\frac{R_H}{r}\right)^{2(d-2)}
\frac{\sinh^2(\alpha_n)\sinh^2(\alpha_w)}{\Delta(\alpha_n)\Delta(\alpha_w)}\right], \label{conf}
\eea
with $\varphi(r)$ still given by (\ref{fr2}).

For later purposes it will useful to have $f=g,$ at least to order $\lambda=0.$ This can be achieved with a conformal transformation which changes the frame, like we have seen in section \ref{diffr}, defining a new metric
\be
g_{\mu\nu}^I = e^{-2\phi} g_{\mu\nu}^S. \label{frs}
\ee
This way we get a solution of the form
\bea
f(r) &=& f_0^I(r) \left(1+ \frac{\lambda}{R_H^2} f_c(r) \right), \, \, g(r)= f_0^I(r) \left(1+ \frac{\lambda}{R_H^2} g_c(r) \right), \label{fcgc} \\
f_0^I&=&\frac{f_0^T}{\sqrt{\Delta(\alpha_n)\Delta(\alpha_w)}}, \label{fcgi}
\eea
$f_0^T$ being given by (\ref{tangher}). $f_c, g_c$ are given by
\bea
f_c^I(r) &=& \frac{1}{2 \Delta(\alpha_n)\Delta(\alpha_w)} \Big(-4 \Delta(\alpha_n)\Delta(\alpha_w) \varphi(r) + 2\left(2-f_0^T\right) \left(\Delta(\alpha_n) \sinh^2(\alpha_w) + \Delta(\alpha_w) \sinh^2(\alpha_n)\right)\mu(r) \nonumber\\
&+& 4 \left. \left(1- \left(\frac{R_H}{r}\right)^{2(d-3)} \sinh^2(\alpha_w) \sinh^2(\alpha_n) \right) \mu(r) + (d-3)^2 f_0^T \left(\frac{R_H}{r}\right)^{2(d-2)} \sinh^2(\alpha_w) \sinh^2(\alpha_n) \right), \nonumber\\
g_c^I(r)&=&\frac{1}{2 \Delta(\alpha_n)\Delta(\alpha_w)} \Big( 2 \left(\Delta(\alpha_n) \sinh^2(\alpha_w) + \Delta(\alpha_w) \sinh^2(\alpha_n)\right)\mu(r) f_0^T \nonumber\\
&+& \left. (d-3)^2 f_0^T \left(\frac{R_H}{r}\right)^{2(d-2)} \sinh^2(\alpha_w) \sinh^2(\alpha_n)+ 4 \Delta(\alpha_n)\Delta(\alpha_w) \left(\varphi(r)-\epsilon(r)\right) \right). \label{fcgorb}
\eea
Since, as discussed also in section \ref{diffr}, the analysis of the stability under tensorial perturbations is independent of the chosen frame, this is the form of the metric we will take.

\subsubsection{Study of the stability}
\indent

In order to prove the stability of this solution, we need to show the positivity of (\ref{qf}), as we have seen. Again it is simpler to split this expression in its "classical" and $\lambda$--corrected parts: $\frac{Q}{F} = \left.\frac{Q}{F}\right|_0 + \lambda \left.\frac{Q}{F}\right|_1,$ with $\left.\frac{Q}{F}\right|_1$ being evaluated using the $\lambda=0$ parts of the corresponding functions. For this concrete solution,
\bea
r^2 \sqrt{fg} \left.\frac{Q}{F}\right|_0 &=& \frac{\ell \left( \ell + d - 3 \right)}{r^2} f + \frac{(g-f)f'}{r} \label{qf0} \\
r^4 \sqrt{fg} \left.\frac{Q}{F}\right|_1 &=& \frac{2}{r^2} \ell \left( \ell + d - 3 \right) f_0^I \left[ 2 (1-f_0^I) + r (f_0^I)' \right].
\eea
$f_0^I(r)$ vanishes at $r=R_H,$ and from there it grows monotonically to the asymptotic value 1. Therefore one has both $(f_0^I)'>0$ and $1-f_0^I>0$ and therefore, in the range we are interested, $\left.\frac{Q}{F}\right|_1>0.$

$\left.\frac{Q}{F}\right|_0$ given by (\ref{qf0}) must be computed with the full, $\lambda$ corrected metric. In this case, from (\ref{fcgc}) and (\ref{fcgorb}) we get
\bea
g-f &=& \frac{2 \lambda f_0^T(r)}{\left(\Delta(\alpha_w) \Delta(\alpha_n) \right)^{\frac{3}{2}}}
\Big[ \Delta(\alpha_w) \Delta(\alpha_n) \left(2\varphi(r)-\epsilon(r)\right) + \left(\left(\frac{R_H}{r}\right)^{2(d-3)} \sinh^2(\alpha_w) \sinh^2(\alpha_n) -1 \right) \mu(r) \nonumber\\
&-& \left(\frac{R_H}{r}\right)^{d-3} \left(\sinh^2(\alpha_n) \Delta(\alpha_w) + \sinh^2(\alpha_w) \Delta(\alpha_n) \right) \mu(r) \Big]
\eea
which, using the definitions (\ref{mu}), (\ref{delta}) and (\ref{fcgi}), may be simplified to
\be
g-f = 2 \lambda \frac{f_0^T(r)}{\sqrt{\Delta(\alpha_w) \Delta(\alpha_n)}}
\Big[2\varphi(r)-\epsilon(r)- \mu(r)\Big] = \frac{4}{d-2} \lambda f_0^I(r)
\left[\left(d-3\right)\varphi(r)+r \varphi'(r)\right].\label{gmf}
\ee

In any other frame but the one we chose by (\ref{frs}), this expression would be much more complicated.

$\Delta(\alpha_w),\, \Delta(\alpha_n)$ are strictly positive functions, as one can see from the definition (\ref{delta}), and so is therefore $f_0^I(r)$. Therefore in order to analyze the positivity of (\ref{gmf}) one needs to concentrate only on the factor inside brackets, $\left(d-3\right)\varphi(r)+r \varphi'(r).$ One can already anticipate that the final result for the stability will not depend on the magnitude of the charges/boost parameters $\alpha_w, \alpha_n,$ a quite remarkable fact.

As we previously mentioned, the dilaton solution $\varphi(r)$ is negative, but its derivative $\varphi'(r)$ is positive: $\varphi(r)$ have a negative value at the horizon and grow up to 0 at asymptotic infinity. The difference $g-f$ is therefore a sum of a positive and a negative term; now, does this mean that $g-f$ does not have a fixed sign, i.e. its sign may change for different values of $r?$ Or is there any dominant term in all the range $r>R_H$, such that the sign of $g-f$ does not change in such range? If this is the case, then which is that sign?

Close to the horizon we have
\bea
g-f &=& -\frac{(d-3)(d-2)}{2 \cosh(\alpha_n) \cosh(\alpha_w)} \frac{\lambda}{R_H^2} \nonumber \\
&\times& \left(\frac{8}{d-1} +d^2 -6d +1 +2(d-3) \left(\psi^{(0)}\left(\frac{2}{d-3}\right) + \gamma \right)\right) \frac{r-R_H}{R_H} + {\mathcal{O}} \left( \left( r-R_H \right)^2 \right)
\eea
One can easily show (again, for instance, constructing a table with Mathematica) that, for relevant values of $d,$ one has $\frac{8}{d-1} +d^2 -6d +1 +2(d-3) \left(\psi^{(0)}\left(\frac{2}{d-3}\right) + \gamma \right) <0$ and, therefore, in some neighborhood of $R_H$ (but of course with $r>R_H$) one has $g-f>0$ (for $r=R_H$ evidently $g=f$).

For large $r$ we obtain the following asymptotic expansion:
\be
g-f = \frac{d-3}{2} \frac{\lambda}{R_H^2} \left(\frac{R_H}{r}\right)^{2(d-3)} \left[d-2 - (d-1) \left(\frac{R_H}{r}\right)^2 \right] + {\mathcal{O}} \left( \left(\frac{R_H}{r}\right)^{2d-3} \right) .
\ee
In the limit $r\rightarrow\infty$ one has $g=f=1,$ i.e. $g-f=0.$ For large $r$ we see that the leading correction is positive: one also has $g-f>0,$ as one had close to the horizon. The same conclusions are reached when one restricts themselves to $\left(d-3\right)\varphi(r)+r \varphi'(r):$ positive at the horizon and vanishing at infinity. It remains to analyze the behavior of $\left(d-3\right)\varphi(r)+r \varphi'(r)$ in the intermediate range (neither close to $R_H$ nor asymptotically). Rather we analyze its derivative:
\be
\left(\left(d-3\right)\varphi(r)+r \varphi'(r)\right)'=-\frac{(d-3)(d-2)^2}{4r} \left(\frac{R_H}{r}\right)^{2d-6} \frac{d-3-(d-1) \left(\frac{R_H}{r}\right)^{2} +2\left(\frac{R_H}{r}\right)^{d-1}}{\left(1-\left(\frac{R_H}{r}\right)^{d-3}\right)^2}. \label{gmflinha}
\ee
One can easily check that, for relevant values of $d$, and for $r>R_H,$ $d-3-(d-1) \left(\frac{R_H}{r}\right)^{2} +2\left(\frac{R_H}{r}\right)^{d-1}>0.$ The other factors in (\ref{gmflinha}) are clearly positive, except for the overall minus sign. Therefore one has $\left(\left(d-3\right)\varphi(r)+r \varphi'(r)\right)'<0$ for $r>R_H$ (in particular this derivative has no zeroes). This means $\left(d-3\right)\varphi(r)+r \varphi'(r)$ is a positive function which decreases to zero asymptotically. The behavior of $g-f$ is the same.

Since for this solution $g-f$ is of order $\lambda$ but it vanishes at the classical level, $f'$ in $\left.\frac{Q}{F}\right|_0$ must be computed with the classical metric, exactly from the same argument we gave for the dilatonic solution (\ref{fnew}) (the $\lambda$--correction of the metric to $f'$ would only contribute at order $\lambda^2$.) As we have mentioned, at order $\lambda=0$ we have $(f_0^I)'>0;$ from our previous result, this way, to order $\lambda,$ $(g-f) f' >0.$

Since, by definition, $f(R_H)=0$ and $f>0$ for $r>R_H,$ we conclude that $\left.\frac{Q}{F}\right|_0$  given by (\ref{qf0}) is indeed positive for $r>R_H.$ The same is valid for $\frac{Q}{F}$ given by (\ref{qf}). This way, the double--charged $d$--dimensional black hole solution (\ref{fcgc}) is indeed stable under tensorial perturbations, for any value of the magnitude of the charges/boost parameters $\alpha_w, \alpha_n$.

\section{Discussion and Future Directions}
\indent

We have computed the tensorial perturbations for the most general spherically symmetric metric in $d$ dimensions. We applied it to the field equations resulting from a string effective action with $\R^2$ corrections. We have shown that the master equation for the perturbation variable is of second order, regardless of the presence of the higher order terms in the lagrangian and in the field equations. We have also obtained the corresponding $\a$ corrections to the potential for these perturbations. From these results we have studied the stability of two different $d$--dimensional black hole solutions with $\R^2$ corrections in string theory. In both cases we concluded that they were stable under such perturbations.

These results are to be compared with the corresponding ones in Lovelock theory. As we have mentioned in the introduction, for tensorial perturbations of solutions to these theories several instabilities have been found \cite{dg04,dg05a,Takahashi:2009dz,Takahashi:2010ye,Takahashi:2010gz}, depending on the dimensionality of spacetime (the necessary power of curvature needed in order to get a Lovelock theory also depends on $d$). Even more interesting are two other aspects of Lovelock theories which were found on these articles: the instabilities manifest themselves mainly on shorter scales (for ``small'' black holes), and there are domains of the parameters (the coupling constants of the higher order terms) in which the linear perturbation theory breaks down and is not applicable. The justification for such facts lies on the properties of Lovelock theories we mentioned: they are seen as exact and not effective theories; the dependence of the solutions on the coupling constants goes beyond perturbation theory, i.e. the order at which they appear in the lagrangian does not matter for such dependence, which is often nonlinear. This is the reason for the nonapplicability of the linear perturbation theory in certain domains.

The string--theoretical solutions we have considered are perturbative in $\a$: their dependence on $\a$ is of the same order in which $\a$ appears on the lagrangian (first order). This is why linear perturbation theory is fully applicable to these solutions we have studied, but one must keep in mind that their stability which we have shown is just perturbative. Nothing in our work guarantees that, if one considers higher order corrections in $\a$ to these solutions, an instability does not appear.

There is a lot of work still to be done in studying perturbations and stability of black holes in higher dimensions, already in classical Einstein gravity. With respect to static black holes for which the formalism of Ishibashi and Kodama is applicable, there is a lot to be understood concerning scalar perturbations (for generic horizon topologies) and tensor perturbations (for non--maximally symmetric black holes). For a recent review see \cite{Ishibashi:2011ws}. There is even more to be done concerning the perturbations and stability of non--static black holes.

With respect to perturbative string--theoretical black holes, the main question one can ask is the following: do string $\a$ corrections preserve the stability properties of the corresponding classical black hole solutions, or do they introduce a new behavior? The examples we have analyzed in this article provide some evidence in favor of the first answer, but they represent by far an extremely limited range of solutions for more definite conclusions to be drawn. Still a lot of work remains to be done, with other string--corrected solutions but even also with those solutions we have taken here, considering other kinds of metric perturbations (vector and scalar).

Knowing the master equation and potential for tensor--type gravitational perturbations for these solutions, as we do, another study that can be made is the determination of the corresponding spectrum of quasinormal modes, including the leading $\a$ corrections. We leave this for a future work.

\section*{Acknowledgments}
The author wishes to acknowledge useful discussions with Akihiro Ishibashi and Jos\'e Sande Lemos. This work has been supported by CMAT - U.Minho through the FCT Pluriannual Funding Program, and by FCT and FEDER through CERN/FP/123609/2011 project.


\bibliographystyle{plain}

\begin{thebibliography}{10}

\bibitem{iks00}
A.~Ishibashi, H.~Kodama and O.~Seto,
\textit{Brane World Cosmology: Gauge Invariant Formalism For Perturbation},
Phys.\ Rev.\ \textbf{D62} (2000) 064022,
\texttt{[hep-th/0004160]}.

\bibitem{ik03a}
A.~Ishibashi and H.~Kodama,
\textit{A Master Equation for Gravitational Perturbations of Maximally Symmetric Black Holes in Higher Dimensions},
Prog.\ Theor.\ Phys.\ \textbf{110} (2003) 701,
\texttt{[hep-th/0305147]}.

\bibitem{ik03c}
A.~Ishibashi and H.~Kodama,
\textit{Master Equations for Perturbations of Generalized Static Black Holes with Charge in Higher Dimensions},
Prog.\ Theor.\ Phys.\ \textbf{111} (2004) 29,
\texttt{[hep-th/0308128]}.

\bibitem{ik03b}
A.~Ishibashi and H.~Kodama, \textit{Stability of Higher Dimensional Schwarzschild Black Holes},
Prog.\ Theor.\ Phys.\ \textbf{110} (2003) 901,
\texttt{[hep-th/0305185]}.

\bibitem{ik03d}
A.~Ishibashi and H.~Kodama, \textit{Stability of Generalized Static Black Holes in Higher Dimensions},
\texttt{[gr-qc/0312012]}.

\bibitem{Konoplya:2008rq}
R.~A.~Konoplya and A.~Zhidenko, \emph{Stability of higher dimensional Reissner--Nordstr\"om--anti--de Sitter black holes}, Phys.\ Rev.\ {\bf D78} (2008) 104017 [arXiv:0809.2048 [hep-th]].

\bibitem{Konoplya:2008au}
R.~A.~Konoplya and A.~Zhidenko, \emph{Instability of higher dimensional charged black holes in the de-Sitter world}, Phys.\ Rev.\ Lett.\ {\bf 103} (2009) 161101 [arXiv:0809.2822 [hep-th]].

\bibitem{Kunduri:2006qa}
H.~K.~Kunduri, J.~Lucietti and H.~S.~Reall, \emph{Gravitational perturbations of higher dimensional rotating black holes: Tensor perturbations}, Phys.\ Rev.\ {\bf D74} (2006) 084021 [hep-th/0606076].

\bibitem{Kodama:2009rq}
H.~Kodama, R.~A.~Konoplya and A.~Zhidenko, \emph{Gravitational instability of simply rotating AdS black holes in higher dimensions}, Phys.\ Rev.\ {\bf D79} (2009) 044003 [arXiv:0812.0445 [hep-th]].

\bibitem{Dias:2009iu}
O.~J.~C.~Dias, P.~Figueras, R.~Monteiro, J.~E.~Santos and R.~Emparan, \emph{Instability and new phases of higher-dimensional rotating black holes}, Phys.\ Rev.\ {\bf D80} (2009) 111701 [arXiv:0907.2248 [hep-th]].

\bibitem{Dias:2010eu}
O.~J.~C.~Dias, P.~Figueras, R.~Monteiro, H.~S.~Reall and J.~E.~Santos, \emph{An instability of higher--dimensional rotating black holes}, JHEP {\bf 1005} (2010) 076 [arXiv:1001.4527 [hep-th]]

\bibitem{Dias:2010maa}
O.~J.~C.~Dias, P.~Figueras, R.~Monteiro and J.~E.~Santos, \emph{Ultraspinning instability of rotating black holes}, Phys.\ Rev.\ {\bf D82} (2010) 104025 [arXiv:1006.1904 [hep-th]].

\bibitem{Dias:2010gk}
O.~J.~C.~Dias, P.~Figueras, R.~Monteiro and J.~E.~Santos, \emph{Ultraspinning instability of anti--de Sitter black holes}, JHEP {\bf 1012} (2010) 067 [arXiv:1011.0996 [hep-th]].

\bibitem{dg04}
G.~Dotti and R.~J.~Gleiser,
\textit{Gravitational Instability of Einstein--Gauss--Bonnet Black Holes under Tensor Mode Perturbations}, Class.\ Quant.\ Grav.\ \textbf{22} (2005) L1, \texttt{[gr-qc/0409005]}.

\bibitem{dg05a}
G.~Dotti and R.~J.~Gleiser, \textit{Linear Stability of Einstein--Gauss--Bonnet Static Spacetimes, Part I: Tensor Perturbations}, Phys.\ Rev.\ \textbf{D72} (2005) 044018, \texttt{[gr-qc/0503117]}.

\bibitem{Takahashi:2009dz}
T.~Takahashi and J.~Soda, \textit{Stability of Lovelock Black Holes under Tensor Perturbations}, Phys.\ Rev.\ {\bf D79} (2009) 104025 [arXiv:0902.2921 [gr-qc]].

\bibitem{Takahashi:2010ye}
T.~Takahashi and J.~Soda, \textit{Master Equations for Gravitational Perturbations of Static Lovelock Black Holes in Higher Dimensions}, Prog.\ Theor.\ Phys.\ {\bf 124} (2010) 911 [arXiv:1008.1385 [gr-qc]].

\bibitem{Takahashi:2010gz}
T.~Takahashi and J.~Soda, \textit{Catastrophic Instability of Small Lovelock Black Holes}, Prog.\ Theor.\ Phys.\ {\bf 124} (2010) 711 [arXiv:1008.1618 [gr-qc]].

\bibitem{Camanho:2009vw}
X.~O.~Camanho and J.~D.~Edelstein, \textit{Causality constraints in AdS/CFT from conformal collider physics and Gauss-Bonnet gravity}, JHEP {\bf 1004} (2010) 007 [arXiv:0911.3160 [hep-th]].

\bibitem{Camanho:2009hu}
X.~O.~Camanho and J.~D.~Edelstein, \textit{Causality in AdS/CFT and Lovelock theory}, JHEP {\bf 1006} (2010) 099 [arXiv:0912.1944 [hep-th]].

\bibitem{Camanho:2010ru}
X.~O.~Camanho, J.~D.~Edelstein and M.~F.~Paulos, \textit{Lovelock theories, holography and the fate of the viscosity bound}, JHEP {\bf 1105} (2011) 127 [arXiv:1010.1682 [hep-th]].

\bibitem{Moura:2006pz}
F.~Moura, R.~Schiappa, \textit{Higher-derivative corrected black holes: Perturbative stability and absorption cross-section in heterotic string theory}, Class.\ Quant.\ Grav.\ {\bf 24} (2007) 361 [hep-th/0605001].

\bibitem{Higuchi:1986wu}
A.~Higuchi, \textit{Symmetric Tensor Spherical Harmonics On The N--Sphere And Their Application To The De Sitter Group SO(N,1)}, J.\ Math.\ Phys.\  {\bf 28 } (1987)  1553.

\bibitem{Moura:2009it}
F.~Moura, \textit{String--corrected dilatonic black holes in d dimensions}, Phys.\ Rev.\ {\bf D83} (2011) 044002
[arXiv:0912.3051 [hep-th]].

\bibitem{cmp89}
C.~G.~Callan, R.~C.~Myers and M.~J.~Perry, \textit{Black Holes in String Theory}, Nucl.\ Phys.\ \textbf{B311} (1989) 673.

\bibitem{Tangherlini:1963bw}
F.~R.~Tangherlini, \textit{Schwarzschild field in n dimensions and the dimensionality of space problem},
Nuovo Cim.\ {\bf 27} (1963) 636.

\bibitem{Giveon:2009da}
A.~Giveon, D.~Gorbonos and M.~Stern, \textit{Fundamental Strings and Higher Derivative Corrections to d--Dimensional Black Holes}, JHEP {\bf 1002} (2010) 012 [arXiv:0909.5264 [hep-th]].

\bibitem{Ishibashi:2011ws}
A.~Ishibashi and H.~Kodama, \emph{Perturbations and Stability of Static Black Holes in Higher Dimensions}, Prog.\ Theor.\ Phys.\ Suppl.\ {\bf 189} (2011) 165 [arXiv:1103.6148 [hep-th]].
\end{thebibliography}

\end{document}